\documentclass[preprint]{aastex}
\def\simgr{\,\hbox{\hbox{$ > $}\kern -0.8em \lower 1.0ex\hbox{$\sim$}}\,}
\def\simle{\,\hbox{\hbox{$ < $}\kern -0.8em \lower 1.0ex\hbox{$\sim$}}\,}
\shortauthors{SHEETS ET AL.}
\shorttitle{SPECTROSCOPY OF NINE CVs}
\begin{document}
\title{Spectroscopy of Nine Cataclysmic Variable Stars
\footnote{Based on observations obtained at the MDM Observatory, operated by
Dartmouth College, Columbia University, Ohio State University, Ohio University, and
the University of Michigan.}
}

\author{Holly A. Sheets, John R. Thorstensen, 
Christopher J. Peters, and Ann B. Kapusta}
\affil{Department of Physics and Astronomy\\
6127 Wilder Laboratory, Dartmouth College\\
Hanover, NH 03755-3528;\\
h.a.sheets@dartmouth.edu}
\author{Cynthia J. Taylor}
\affil{The Lawrenceville School\\
P.O. Box 6008, Lawrenceville, NJ 08648}

\begin{abstract}
We present optical spectroscopy of nine cataclysmic binary stars,
mostly dwarf novae, obtained primarily to  
determine orbital periods $P_{\rm orb}$.
The stars and their periods are
LX And, 0.1509743(5) d;
CZ Aql, 0.2005(6) d;
LU Cam, 0.1499686(4) d;
GZ Cnc, 0.0881(4) d;
V632 Cyg, 0.06377(8) d;
V1006 Cyg, 0.09903(9) d;
BF Eri, 0.2708804(4) d;
BI Ori, 0.1915(5) d; and
FO Per, for which $P_{\rm orb}$ is either 0.1467(4) or 0.1719(5) d.

Several of the stars proved to be especially interesting.
In BF Eri, we detect the absorption spectrum of a secondary star
of spectral type K3 $\pm 1$ subclass, which leads to a distance
estimate of $\sim 1$ kpc. However, BF Eri has a large proper motion
($\sim 100$ mas yr$^{-1}$), and we have a preliminary 
parallax measurement that confirms the large proper motion
and yields only an upper limit for the parallax.  BF Eri's
space velocity is evidently large, and it appears to belong 
to the halo population.  In CZ Aql, the emission lines have strong 
wings that move with large velocity amplitude, suggesting a 
magnetically-channeled accretion
flow.  The orbital period of V1006 Cyg places it squarely
within the 2- to 3-hour `gap' in the distribution of 
cataclysmic binary orbital periods. 
\end{abstract}


\section{Introduction}

Cataclysmic variables (CVs) are binary star systems in which the secondary,
usually a late-type main sequence star, fills its Roche lobe and loses mass 
to the white dwarf primary \citep{warner}.  CVs are long-lived systems that
are stable against mass transfer, so the mass transfer must be
driven by gradual changes in the orbit, or in the secondary star, or both.  
It is commonly believed that the evolution of most CVs is driven by the 
slow loss of angular momentum from the orbit, most likely through magnetic braking
of the co-rotating secondary star, at least at longer orbital
periods $P_{\rm orb}$ where gravitational radiation is ineffective 
(\citealt{andronov} give a recent discussion). 
The loss of angular momentum constricts the Roche critical lobe
around the secondary and causes the system to transfer mass as it evolves
toward shorter $P_{\rm orb}$.  In this scenario, 
$P_{\rm orb}$ serves as a proxy measurement for the system's 
evolutionary state.  Correct and complete orbital period measurements
are fundamental to any accurate theory of CV evolution.
Given the usefulness of $P_{\rm orb}$, it is fortunate that it 
can usually be measured accurately and precisely.

This paper presents optical spectroscopy of the nine CVs listed in 
Table \ref{tab:synthmags}.
We took these observations mostly for the purpose of finding orbital periods
using radial velocities (none of these systems are known to eclipse). 
The long cumulative exposures also allowed us to look for any unusual features.
The {\it Catalog and Atlas of Cataclysmic Variables Archival Edition} \citep{Downes}
\footnote{Available at http://archive.stsci.edu/prepds/cvcat/index.html; this had
been called the {\it Living Edition} until its author retired and ceased 
updates.}
lists seven of the stars as dwarf novae, one as either a dwarf nova or a DQ Her
star, and one simply as a cataclysmic, possibly a dwarf nova similar to 
U Gem or SS Cygni (type UGSS).  
Except for CZ Aql, for which we confirm a 4.8-hour candidate period
suggested by \citet{ciess},
all of these objects lacked published orbital periods when we began working
on them.  Subsequently \citet{tappert} found $P_{\rm orb} = 0.0883$ d for
GZ Cancri; we had communicated our advance findings to these authors so 
they could disambiguate their period determination.  

\section{Observations, Reductions, and Analysis}

\subsection{Observations}

All our spectra were taken at the MDM Observatory on Kitt Peak,
Arizona, using either the 1.3m McGraw-Hill telescope or the 2.4m
Hiltner telescope.  The earliest observations we report here are from 1995,
and the latest were obtained 2007 January.  Table \ref{tab:obsjournal} 
gives a journal of the observations.

At the 1.3m we used the Mark III spectrograph and a SITe
$1024\times1024$ CCD detector.  The spectral resolution is 5.0 \AA, covering a 
range of either 4480 to 6760 \AA\ with 2.2 \AA\ pixel$^{-1}$ for the 2001 
December BF Eri 
data, or 4646 to 6970 \AA\ with 2.3 \AA\ pixel$^{-1}$ for the remaining data.  
The 2.4m spectra, except for those of FO Per, were obtained with the 
modular spectrograph and a SITe $2048^2$ CCD 
detector, with 2.0 \AA\ pixel$^{-1}$, over a range 
of 4210 to 7500 \AA\, and with a spectral resolution of 3.5 \AA.  The
relatively small number of 2.4 m spectra of FO Per were
taken with a LORAL $2048^{2}$-pixel detector, and cover
from 4285 to 6870 \AA\ at 1.25 \AA\ pixel$^{-1}$.

\subsection{Reductions}

For the most part we reduced the spectra using 
standard IRAF\footnote{IRAF is distributed by the National Optical Astronomy
Observatories.} procedures.   The wavelength calibration was based on 
exposures of Hg, Ne, and Xe lamps.  Prior to 2003 we took
lamp exposures through the night and whenever the telescope was moved.
For the 2.4m data from 2003 to the present, we used lamp exposures taken 
in twilight to find the shape of the 
pixel-to-wavelength relation, and set the zero point individually for each 
nighttime exposure using the OI $\lambda 5577$ night-sky feature. 
The apparent velocity of the telluric OH emission bands at the far red end of the spectrum,
found with a cross-correlation routine,
provided a check; although these are far from the feature used to set the 
zero point, their apparent velocity typically remain within 10 km s$^{-1}$ of
zero.  Because of the increased efficiency of this technique, we attempted 
to use it at the 1.3m telescope also, during the 2004 June/July observing run.
For unknown reasons the results were unsatisfactory.  
To salvage the H$\alpha$ emission velocities from that run, we determined a 
correction by cross-correlating the
night-sky emission features in the 6200-6625 \AA\ range 
with a well-calibrated night-sky spectrum obtained with a similar instrument.
The correction was  
calculated for each individual spectrum and then applied to each measured
velocity, and it did reduce the scatter somewhat, evidently because the
wavelength range used includes the H$\alpha$ emission line for which we
measured velocities.

On all our runs, we observed flux standards during twilight when the
sky was clear, and applied the resulting calibration to the data.
The reproducibility of these observations suggests that our fluxes
are typically accurate to $\pm 20$ per cent.
We also took short exposures of bright O and B stars in twilight to map the 
telluric absorption features and divide them out approximately from our
program object spectra.
Before flux calibration, we divided our program star spectra by a mean
hot-star continuum, in order to remove the bulk of the response variation.
Table \ref{tab:synthmags} lists $V$ magnitudes synthesized from our mean
spectra, using the IRAF {\it sbands} task and the passband tabulated
by \citet{bessell}; clouds, losses at the slit, and calibration errors
make these uncertain by a few tenths of a magnitude, but they do give
a rough indication of the brightness of each system at the time of our
observation.

\subsection{Analysis}

\begin{figure}
\epsscale{0.9}
\plotone{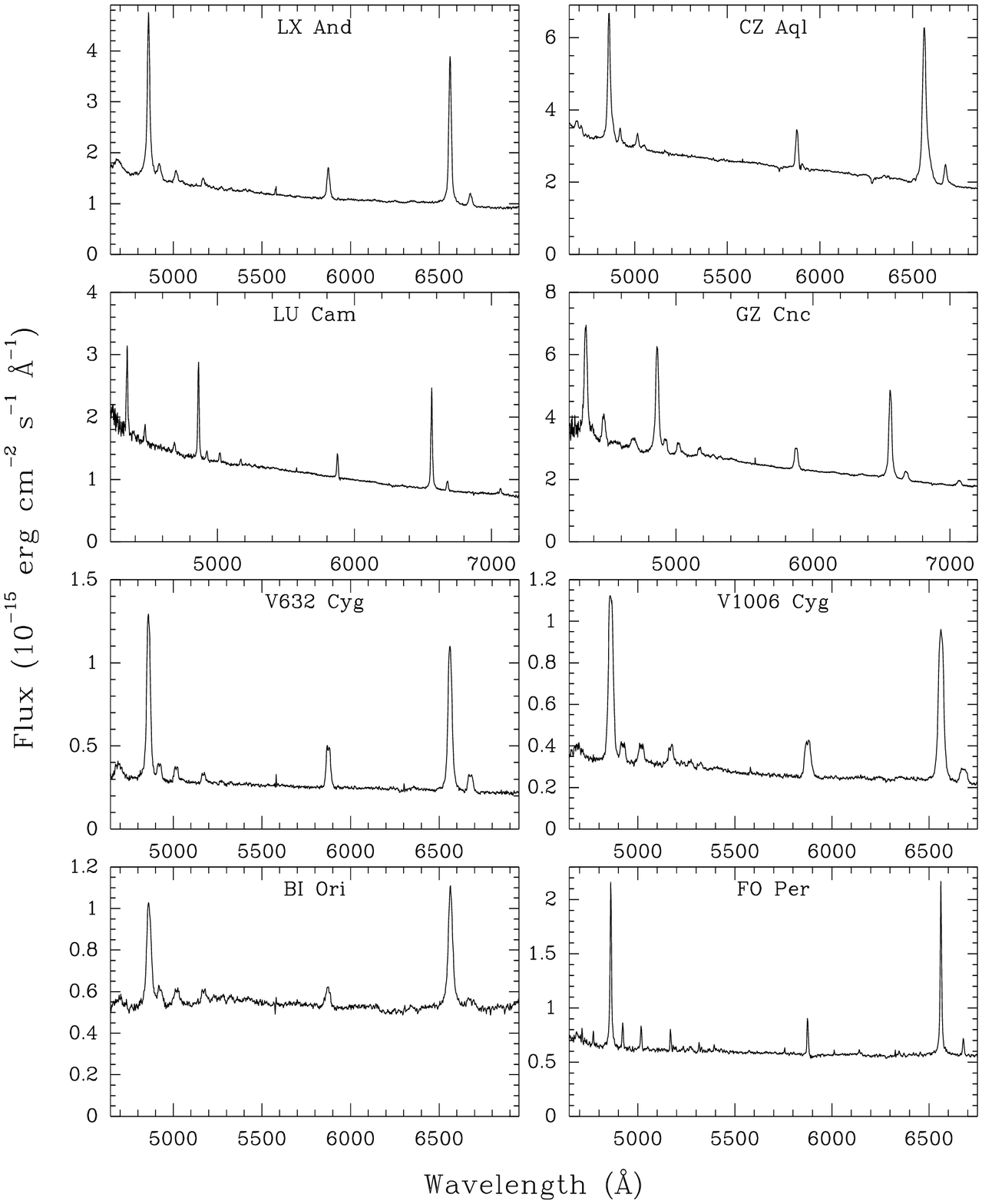}
\caption{Plots of the average flux-calibrated spectra for eight of the stars studied here. 
The weak features seen near $\lambda 5577$ are artifacts caused by imperfect subtraction of the
strong [OI] night-sky emission.}
\label{fig:specplot}
\end{figure}

Except for a few spectra taken in outburst (which show weak emission or
absorption on a strong continuum), all of the stars show the prominent
emission lines.  Figs.~\ref{fig:specplot} and  ~\ref{fig:bfsppl} show
averaged spectra, and Table~\ref{tab:quie} gives the 
equivalent width and FWHM of each line measured for each star from its averaged
spectrum.

Two stars, BF Eri and BI Ori, showed the spectral features of a late-type
star.  
To quantify the secondary contribution
in these objects, we began by preparing averaged flux-calibrated
spectra (in BF Eri's case the secondary's radial velocity curve
was measurable, so we shifted the individual spectra to the 
secondary's rest frame before averaging).  Over time we have 
used the 2.4 m and modular spectrograph to collect
spectra of K and M stars classified by \citet{keenan} or 
\citet{boeshaar}.  The wavelength coverage and spectral
resolution of these data are similar to the 1.3m data.
We applied a range of scaling factors to the
library spectra, subtracted them from the averaged spectra,
and examined the results by eye to estimate a range of 
spectral types and scaling factors giving acceptable
cancellation of the late-type features.

We use the spectral type and secondary flux to 
estimate the distance in the following
manner.  We begin by finding the surface brightness of the 
secondary star in $V$, on the assumption that 
the surface brightness is similar to that of main-sequence
stars of the same spectral type; the Barnes-Evans relation
for late-type stars is discussed by \citet{beuermann06}.
Combining the known 
$P_{\rm orb}$ with the assumption that the secondary
fills its Roche critical lobe yields the secondary's 
radius $R_2$ as a function of its mass, $M_2$. 
In the relevant range of mass ratio, $R_2 \propto M_2^{1/3}$, approximately,
and the dependence on $M_1$ is weak enough to ignore. 
We generally do not know $M_2$, so we 
guess at a generous allowable range for this parameter using evolutionary
simulations by \citet{baraffe} as a guideline; the weakness of the dependence
of $R_2$ on $M_2$ means that this (rather questionable) step 
does not dominate the error
budget.  Combining the surface brightness with $R_2$ yields
the absolute magnitude $M_V$.   Subtracting this from the 
apparent magnitude measured for the secondary star 
gives a 
distance modulus.  The reddening maps of \citet{schlegel} then
can be used to estimate the extinction.  Note carefully that
we do not assume that the secondary
is a `normal' main-sequence star; we assume only that the 
secondary's surface brightness is similar to field stars of the same 
spectral type.  The normalization of the secondary's contribution
also depends on the assumption that the spectral features used
to judge the subtraction are similar in strength to 
those of a normal star.

As noted earlier, the immediate aim of our observations was 
to find orbital periods from radial velocity time series data.
The H$\alpha$ emission line is usually the strongest feature, and it 
generally gives good results in dwarf novae.  All the 
emission-line velocities reported here are of H$\alpha$.

We measured radial velocities of H$\alpha$ emission using  
convolution methods described by  \citet{schyo} and \citet{shaf}.  
In this technique one convolves an antisymmetric function with the 
line profile, and takes the zero of the convolution (where the two sides
of the line contribute equally) as the line center.  For the antisymmetric
function with which the spectrum is convolved, we used either  
the derivative of a Gaussian with adjustable width, or positive 
and negative Gaussians of adjustable width offset from each other
by an adjustable separation.
Uncertainties in the convolution velocities are estimated by 
propagating forward
the counting-statistics errors in the individual data channels; in
practice, these are lower limits to the true uncertainties, since the line
profile can vary in ways unrelated to the orbital modulation.
The choice of convolution parameters is dictated by the
shape and width of the line, and in practice the parameters are adjusted
to give the best detection of the orbit.  The 
physical interpretation of CV emission lines is complicated and 
controversial (see, e.g., \citealt{shaf}, \citealt{marsh88},
\citealt{robinson}), but in almost all cases the emission-line periodicity 
accurately reflects $P_{\rm orb}$ (though \citealt{hasitall} 
describe a noteworthy exception to this rule).
A sample of the radial velocities 
for each object are listed in Table~\ref{tab:vels}, while the full tables can
be found online.

One of our systems, BF Eri, has a K-type absorption component in its spectrum.  
We measured velocities of this using the cross-correlation radial velocity package
described by \citet{kurtzmink}, using the region from 5000 to 
6500 \AA , and excluding the region containing the He I $\lambda$5876 
emission line and and the NaD absorption complex.  For a
cross-correlation template spectrum, we used
the a velocity-compensated sum of many observations of IAU velocity
standards taken with the same instrument, as described in \citet{longp03}.

\begin{figure}
\epsscale{0.73}
\plotone{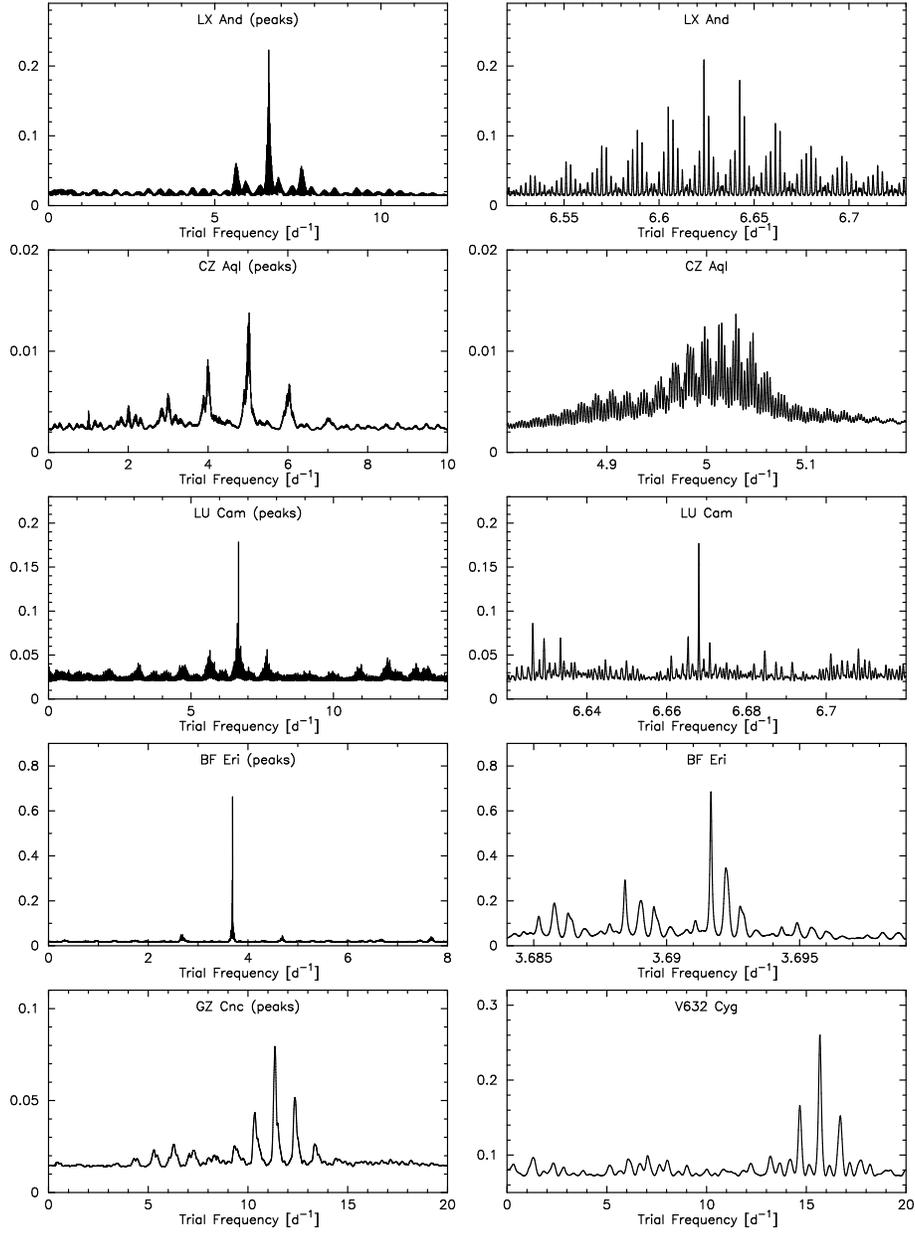}
\caption{Periodigrams for most of the stars studied here.  The vertical
axis in each case is the inverse of chi-square for the least-squares best
fitting sinusoid at each trial frequency.  When data from more that one observing
run are combined, the periodigram can require hundreds of thousands of points
to resolve the fine-scale ringing; in those cases, the curve shown is 
formed by connecting local maxima of the periodogram with straight lines.
In those cases the right-hand panel gives a close-up view of the region
around the highest peak, revealing the alias structure resulting from
different choices of cycle count between the observing runs.
The periodogram of BF Eri is for the absorption-line velocities.
}
\label{fig:pgrm1}
\end{figure}

\begin{figure}
\epsscale{0.72}
\plotone{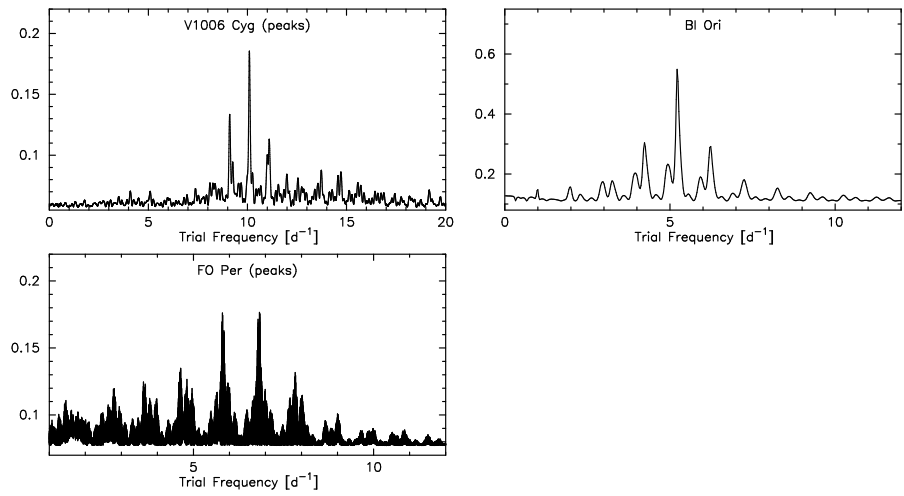}
\caption{Periodigrams for the remainder of the stars, plotted in the same
manner as the previous figure.  Because the choice of daily cycle count for
FO Per remains ambiguous, we have not chosen to enlarge either peak region.  
}
\label{fig:pgrm2}
\end{figure}

\begin{figure}
\epsscale{0.78}
\plotone{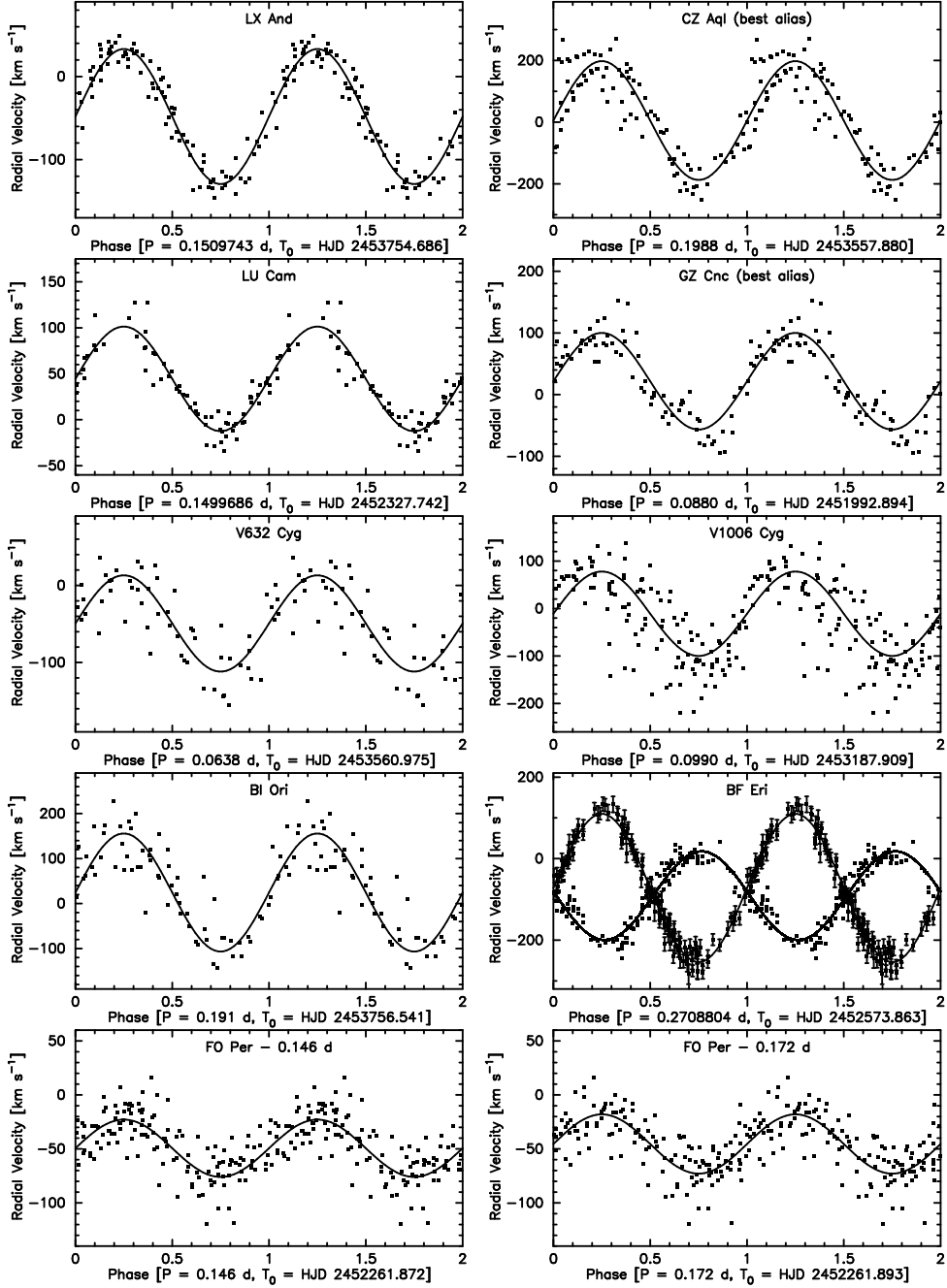}
\caption{Radial velocities plotted as a function of phase using the 
adopted orbital periods.  For CZ Aql, GZ Cnc, and V1006 Cyg, the number
of cycle counts between observing runs is unknown, and the exact period 
chosen to fold the velocities is one of a number of possibilities.
The two plots for FO Per are for different choices of the daily 
cycle count, and each of these in turn is also an arbitrary choice
among many finely-spaced periods.  In BF Eri, both emission and 
absorption velocities are plotted; the absorption velocities are shown
with error bars.}
\label{fig:folpl} 
\end{figure}

We searched for periods in all the velocity time series using the
``residualgram" method \citep{tpst}; the resulting periodograms
are given in Figs.~\ref{fig:pgrm1} and \ref{fig:pgrm2}.  At the best candidate periods we
fitted least-squares sinusoids of the form $v(t) = \gamma + K\sin[2\pi(t-T_0)/P]$.
Fig.~\ref{fig:folpl} shows the velocities folded on the best-fitting 
periods, and Table \ref{tab:param} gives the parameters of these fits.
Because of limitations of the sampling (e.g., the need to observe
only at night from a single site), a single periodicity generally
manifests as a number of alias frequencies.   To assess the
confidence with which we could assert that the strongest alias 
is the true period, we 
used a Monte Carlo test described by \cite{tf85}.

The alias problem can be particularly irksome over longer 
timescales; in this case the uncertain number of cycles elapsed 
between observing runs causes fine-scale ``ringing'' in the 
periodogram.  The individual periods have tiny error bars,
because of the large time span covered, but the ambiguity 
in period means that a realistic error bar -- one that covers
the range of possibilities -- is much larger. 
In those cases, the period uncertainties given in 
Table~\ref{tab:param} are estimated by analyzing data from
the individual observing runs separately.   
When only two observing runs are available, the 
allowable fine-scale frequencies are well-described
by a fitting formula
$$P_{\rm orb}=(t_{2}-t_{1})/n.$$ 
Here $t_{1}$ and $t_{2}$ are the epochs of blue-to-red
velocity crossing observed on the two runs, and $n$ is the 
integer number of cycles that have passed between
$t_{1}$ and $t_{2}$.   The allowed range of $n$ is 
determined from the weighted average of the periods derived 
from separate fits to the two runs' data.  When more 
than two observing runs are available, the situation becomes
more complex.  In some happy cases there are enough overlapping
constraints that only a single, very precise period remains
tenable.  We were able to find such precise periods for LX And, 
LU Cam, and BF Eri.

\section{Notes on Individual Objects}

We discuss the stars in alphabetical order by constellation.

\subsection{LX Andromedae}

LX And was first identified as a variable star (RR V-3) in the Lick RR
Lyrae search \citep{lick82}.  It was classified incorrectly as an RV
Tauri star, and its dwarf nova nature was unrecognized until
the photometric study by \citet{jan00}.  \cite{spec} obtained
spectra of LX And as part of their study of dwarf novae in outburst and
determined the equivalent widths and FWHMs of the Balmer and He II lines.
Our mean spectrum appears typical for a dwarf nova at minimum light.

Because of the large hour-angle span, the radial velocity time series
leaves no doubt about the daily cycle count, which is near 6.6 cycle
d$^{-1}$.  The several observing runs constrain the fine-scale 
period in a more complicated way, but the Monte Carlo test indicates
that a precise period of  0.1509743(5) d is preferred with about 98 per 
cent confidence.  Two other candidate periods separated from
this by 1 cycle per 53.2 d in frequency are much less likely.

\subsection{CZ Aquilae}

Very little has been published on CZ Aql, which is listed in 
the {\it Archival Edition} as a U-Gem dwarf nova.  \citet{ciess} 
included the star
in their spectroscopic study of irregular variables, and noted a
probable 4.8 hour period and emission lines typical of dwarf novae.
Our velocities confirm the suggested 4.8-hour period, but we cannot 
determine a unique cycle count between our observing runs.

While the spectrum superficially resembles that of a dwarf nova, 
a closer look reveals interesting behavior.  Fig.~\ref{fig:trail},
constructed using methods described by \citet{lspeg},   
presents our spectra as a phase-averaged greyscale image.
There is a striking broad component in the stronger Balmer and 
HeI lines that shows a large velocity excursion, with the
red wing of H$\alpha$ reaching to $+3100$ km s$^{-1}$ at phase
0.3 (where phase 0 corresponds to the blue-to-red crossing of the 
line core).  The broad components around H$\beta$ and 
$\lambda 6678$ move in phase with those of H$\alpha$ and 
range from 900 to 2600 and $-2500$ to $-900$ km s$^{-1}$ and 700 to 2100 and 
$-1000$ to $-600$ km s$^{-1}$, respectively.  The wings of $\lambda 5876$ also move
in phase with the others, but the red edge is difficult to follow at its
minimum because of interference from the NaD absorption lines, which
are stationary and hence interstellar.  The
maximum of the red edge is 3200 km s$^{-1}$, while the blue edge ranges from 
$-1500$ to $-600$ km s$^{-1}$.  The blueward wing of all these lines is noticeably
weaker than the redward wing.

Other emission lines present include HeII $\lambda 4686$, HeI $\lambda 4713$ 
and $\lambda 4921$, and, very weakly, FeII $\lambda 5169$.  We also detect 
unidentified emission lines at $\lambda 6344$, as is also seen in LS Peg 
\citep{lspeg}, and at $\lambda 5046$.  The strength of the $\lambda 5780$
diffuse interstellar band \citep{dib} and the NaD lines
suggest that a good deal of interstellar material lies
along the line of sight, and that the luminosity is relatively high.

High-velocity wings reminiscent of the ones seen here have been seen in 
V795 Her \citep{v795her,v795her2},
LS Peg \citep{lspeg},
V533 Her \citep{v533her},
and
RX J1643+34 \citep{rx},
all of which are SW Sex stars.  We do not, however, 
detect another SW Sex characteristic, namely phase-dependent absorption 
in the HeI lines \citep{thorsw}.  The orbital periods of most SW Sex
stars are shorter than 4 hours, so CZ Aql's 4.8-h period would be
unusually long for an SW Sex star.

\subsection{LU Camelopardalis}

\citet{jiang} obtained the first spectrum of this dwarf nova 
in a follow-up study of CV candidates from the ROSAT All Sky Survey.
We found no other published spectroscopic studies.
Our velocities constrain the period to a unique value, 
0.1499685(7) d.  The averaged spectrum shows a rather strong, blue
continuum, which may indicate a state somewhat above true minimum.

\subsection{GZ Cancri}

\citet{jiang} confirmed the cataclysmic nature of GZ Cnc by obtaining
the first spectrum of the object.  \citet{kat02} suggested that this
star, originally labeled as a dwarf nova, could possibly be an intermediate
polar (DQ Her star), based on similarities in its long-term 
photometric behavior to that of
other intermediate polars.  \citet{tappert} conducted a photometric and
spectroscopic study of the system.  Using advance results from the 
present study to help decide the daily cycle count, they found
$P_{\rm orb} = 0.08825(28)$ d,
or 2.118(07) h, placing the system near the lower edge of the so-called
gap in the CV period distribution -- a dearth of systems in the period
range from roughly 2 to 3 hr.  
\citet{tappert} also saw 
characteristics that could indicate an intermediate polar classification, 
but did not claim their evidence was definitive on this point.

Almost all our observations come from two observing runs a year apart.  The 
full set of velocities strongly indicates an orbital frequency near 11.4 
cycle d$^{-1}$, with the Monte Carlo test giving a discriminatory power
greater than 0.99 for the choice of daily cycle count.  However,
the number of cycles between the two observing runs is not determined.
Precise periods that fit the combined data set are given by 
$P = [349.785(3)\ {\rm d}] / n$, where $n$ is the integer number
of cycle counts; $n = 3972 \pm 8$ corresponds to roughly $1$ 
standard deviation.
While our period agrees well with that of \citet{tappert}, 
our data neither support nor disprove the claim that GZ Cnc 
may be an intermediate polar.

\subsection{V632 Cygni}

\citet{v632spec} offer the only published spectrum of this dwarf nova.
They measured the equivalent widths and integrated line fluxes of the
Balmer, HeI, and HeII emission lines and suggested that the orbital period is
likely short based on the very strong Balmer emission.  Our spectrum appears
similar to theirs, and our measured flux level is also nearly the same.
The periodigram in Fig.~\ref{fig:pgrm1} clearly favors an orbital
frequency near 15.7 cycles d$^{-1}$,
with a discriminatory power of 95 per cent and a correctness likelihood
near unity.  
This confirms the suggestion of \citet{v632spec} that the period is rather
short and suggests that it is an SU UMa-type dwarf nova.  

\subsection{V1006 Cygni}

\citet{v1006spec} present the only published spectrum we know of, and
characterized it as a ``textbook example'' of a dwarf nova spectrum.  
They noted a slightly blue continuum with strong Balmer
and He I emission, as well as clear He II $\lambda 4686$ and Fe II emission.
Our spectrum (Fig.~\ref{fig:specplot}) is similar to theirs both in appearance
and normalization, and our line measurements (Table~\ref{tab:quie}) 
are also comparable.

The periodogram (Fig.~\ref{fig:pgrm2}) indicates a frequency near 10.1 cycles
d$^{-1}$, and the Monte Carlo test confirms that the daily cycle count is
securely determined.
Most of our data are from 2004 June, but we returned in 2005 June/July 
to confirm the unusual period indicated in the earlier data.
The periods found by analyzing the two runs separately are consistent 
within their uncertainties.  As with 
GZ Cnc, there are multiple choices for the cycle count between the two
observing runs; the best-fitting periods are given by 
$P = [369.006(4)\ {\rm d}] / n$, where $n = 3726 \pm 4$ corresponds
to 1 standard deviation.  Including a few velocities from other
observing runs suggests that $n$ is slightly larger, perhaps 3728.  In
any case, the period amounts to 2.38 h, which places V1006 Cyg firmly in 
the period gap \citep{warner}, where there is apparently a true
scarcity of dwarf novae \citep{helliergap}.

\subsection{BF Eri}

The first evidence that BF Eridani was a cataclysmic variable came when an {\it
Einstein} X-ray source, 1ES0437-046, was matched to the variable \citep{slew}.
\citet{bfspec} confirmed this match and presented an optical spectrum.
\citet{kat99} and the Variable Star Observers' League in Japan (VSOLJ) 
found photometric variability characteristic of a dwarf nova.  

\begin{figure}
\epsscale{1.0}
\plotone{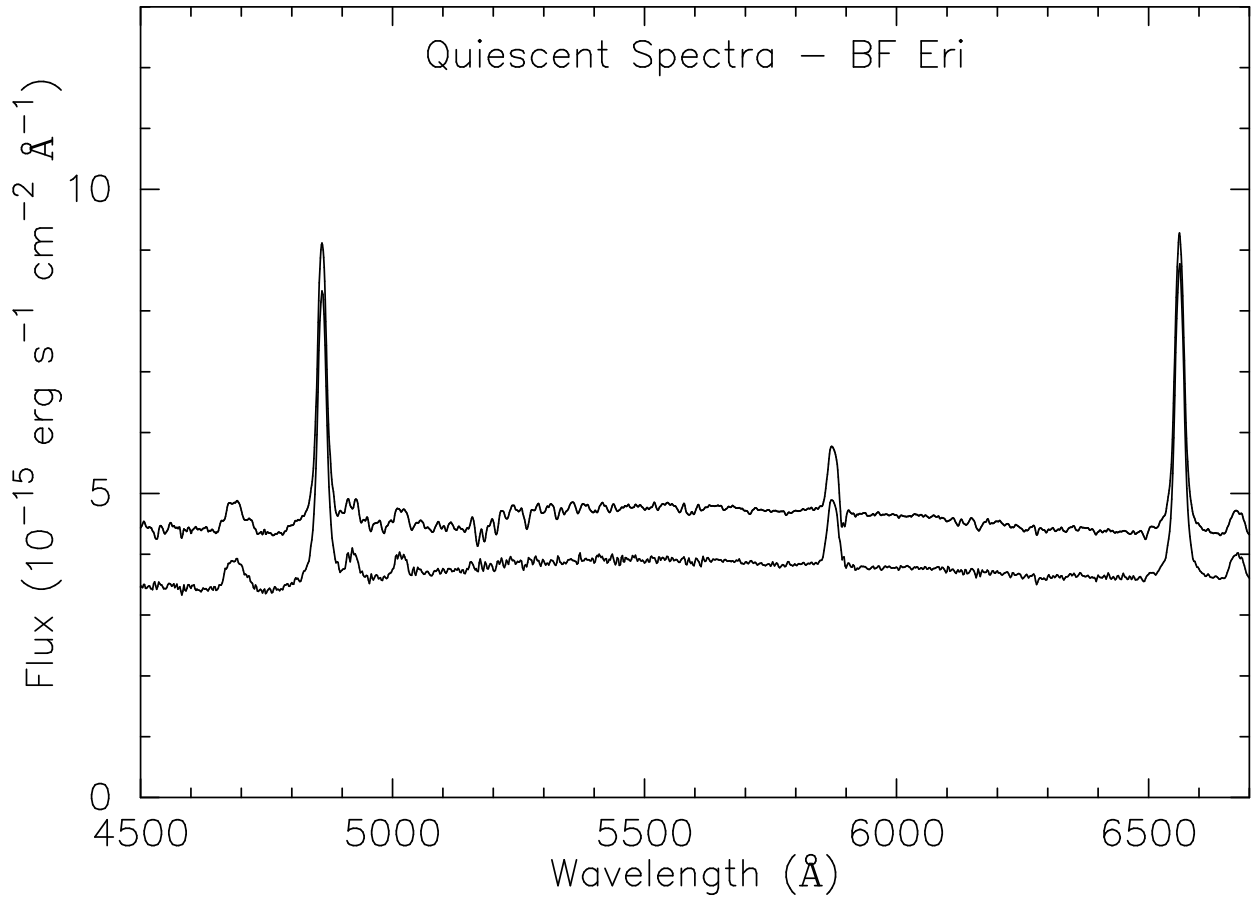}
\caption{Plot of the averaged spectrum of BF Eri (top) and the spectrum after
 scaled late-type (K3V) star has been subtracted (bottom).  The spectra have 
been shifted into a rest frame before averaging and do not include the 2006
March or 2007 January data.}
\label{fig:bfsppl}
\end{figure}

The spectrum of BF Eri (Fig.~\ref{fig:bfsppl}) shows a significant 
contribution from a K star along with the usual dwarf-nova emission lines.
Normally, this suggests that $P_{\rm orb} > 6$ h.  Nearly all our
spectra yielded good cross-correlation radial velocity 
measurements as well as emission-line velocities.  The absorption- and
emission-line velocities independently give a period near
6.50 h (Table~\ref{tab:param}), in accordance with expectation
based on the spectrum.  There is no ambiguity in cycle count 
over the 5-year span of the observations, so the period is
precise to a few parts per million.  Fig.~\ref{fig:trail} shows a 
phase-resolved average of the BF Eri spectra, with the absorption 
spectrum shifting in antiphase to the emission lines.  

If the emission-line velocities faithfully trace the primary's
center-of-mass motion, and the absorption-line velocities 
also trace the secondary's motion, then the two velocity curves 
should be exactly one-half cycle out of phase.
In BF Eri, we find a shift of $0.515 \pm 0.007$ cycles
between the two curves, consistent with 0.5 cycles, so
we feel emboldened to explore the system dynamics.

Masses can only be derived when the 
orbital inclination is known, as in eclipsing systems.
To see if BF Eri might eclipse, we derived differential
magnitudes from images that were taken for astrometry
(discussed below) and plotted them as a function of orbital 
phase.  Some images were taken at the phase at which an eclipse 
would appear, but no evidence for an eclipse was found.  Limits
on the depth and duration of the eclipse are difficult to 
quantify because the data were taken in short bursts in
the presence of strong intrinsic variability, so 
a weak eclipse cannot be ruled out, but the photometry
does suggest that the inclination is not close
to edge-on.

Because the system apparently does not eclipse, we
cannot derive masses; rather, we
find broad constraints on the inclination by assuming
astrophysically reasonable masses for the components. 
Taken at face value,
the velocity amplitudes $K$ imply a mass ratio
$q = M_2 / M_1 = 0.60 \pm 0.03$.  If we arbitrarily
choose a white dwarf mass $M_1 = 0.9$ M$_{\odot}$
(so that $M_2 = 0.53$  M$_{\odot}$), the observed
$K$ velocities imply $i = 50$ degrees.  To find a 
rough lower limit on the inclination, we consider a 
massive white dwarf ($M_1 = 1.2$ M$_{\odot}$) and, 
ignoring the constraint on $q$ for the moment, 
take $M_2 = 0.4$ M$_{\odot}$; this yields $i = 40$ 
degrees.  For a rough upper limit, we assume 
$M_1 = 0.6$ M$_{\odot}$ and $M_2 = 0.4$ M$_{\odot}$,
which gives $i = 67$ degrees.  

The decomposition procedure described earlier yielded a 
spectral type of K3 $\pm1$ subclass; the result of the 
subtraction is shown in Fig.~\ref{fig:bfsppl}.  Using the $V$ passband
tabulated by \citet{bessell} and the IRAF {\it sbands} task,
we find a synthetic $V = 16.9 \pm 0.3$ for the K star's 
contribution.  Taking the range of plausible secondary
star masses to be 0.4 to 0.8 M$_{\odot}$ yields 
$R_2 = 0.7 \pm 0.1$ R$_{\odot}$ at this $P_{\rm orb}$.
Combining this with the surface brightness expected at
this spectral type yields $M_V = 6.8 \pm 0.4$ for the 
secondary.  If there is no significant interstellar 
extinction, we have $m - M = 10.1 \pm 0.5$, or
a distance of approximately $1100 \pm 300$ pc. 
The dust maps of \citet{schlegel} give
a total $E(B-V) = 0.062$ in this direction.
Assuming that BF Eri is beyond the Galactic dust 
and taking $A_V/E(B-V) = 3.3$ gives an 
extinction-corrected $(m - M)_0 = 9.9$, and
a distance estimate of 950 ($+250 ,-200$) pc.  

We can also estimate a distance using the relation
found by \citet{warn87} between $P_{\rm orb}$, $i$, and
the absolute magnitude at maximum light $M_V({\rm max})$.
Using our inclination constraints, the Warner relation
predicts $M_V{\rm max} = 3.9 \pm 0.7$ at this 
orbital period.  The General Catalog of Variable 
Stars \citep{gcvs} lists $m_p = 13.2$ at maximum light; 
taking this to be similar to $V_{\rm max}$ yields 
$m - M = 9.3$, or 9.1 corrected for extinction, which
corresponds to 660 pc. 

Given these distance estimates, it is surprising that BF Eri has a 
very substantial proper motion.  The Lick proper motion survey 
\citep{hanson} gives $[\mu_X, \mu_Y] = [+34, -97]$ mas yr$^{-1}$.
We have begun a series of parallax observations with the
Hiltner 2.4m telescope using the protocols described
by \citet{parallax}; so far we have five epochs from
2005 November and 2007 January.  The proper motion
relative to the background stars is $[\mu_X, \mu_Y] = 
[32, -111]$ mas yr$^{-1}$, and the parallax is not
detected, with a nominal value of 
$1 \pm 2$ mas.  The parallax determination is 
very preliminary, but given the data so far we 
estimate the lower limit on the distance based on 
the astrometry alone to be $\sim 200$ pc.  


At the nominal 950 pc distance derived from the
secondary star, a 100 mas yr$^{-1}$
proper motion corresponds to a transverse velocity
$v_T = 451$ km s$^{-1}$.  This is implausibly large,
so we are left wondering how we might have 
overestimated the distance.  One effect might
be as follows.  Our distance is based on the 
secondary's apparent brightness, and we estimate 
the secondary's contribution to the total light 
by searching for the best cancellation of its
features.  If the secondary's absorption lines
are weaker than those in the spectral-type
standards, we would underestimate the secondary's
contribution.
In our best decomposition, the secondary is 
about 2.2 magnitudes fainter than the total
light in $V$.  Assuming (unrealistically) that 
{\it all} the light is from the secondary 
would therefore decrease the distance modulus by 
2.2 magnitudes, to a distance of 340 pc.

We do not yet have enough information to resolve
the conundrum posed by BF Eri's 
unmistakably large proper motion and its
apparently large distance, but a reasonable 
compromise might be to put it at 
something like 400-500 pc, with an underluminous,
low-metallicity secondary.  The 
cross-correlation velocities of the secondary 
have a zero point determined to $\pm 5$ km s$^{-1}$,
more or less, and give a substantial systemic velocity of 
$-72 \pm 3$ km s$^{-1}$, or $-86$ km s$^{-1}$ in the 
local standard of rest.  If the star is at 450 pc,
its space velocity with respect to the local
standard of rest is $\sim$250 km s$^{-1}$, with Galactic
components $[U,V,W]$ = $[-180,-180,-3]$ km s$^{-1}$, that
is, the velocity is mostly parallel to the Galactic plane 
and lags far behind the rotation of the Galactic disk.
This would put BF Eri on a highly eccentric orbit; these are 
halo-population kinematics (even though the star remains close 
to the plane).  These kinematics would be qualitatively 
consistent with the weak-line conjecture used earlier.

\subsection{BI Ori}

\citet{biori87} published the first quiescent spectrum of BI Orionis, which
showed emission lines typical of dwarf novae. \citet{spec} show a
spectrum in outburst and note the possible presence of weak HeII
$\lambda 4686$ emission.  

Only the 2006 January velocities are extensive enough for period finding;
they give $P_{\rm orb} = 4.6$ hr, with no significant ambiguity in the 
daily cycle count.  The average spectrum shows the usual emission lines;
M-dwarf absorption features are also visible, though the signal-to-noise of the 
individual spectra was not adequate for finding absorption-line velocities.  
Using the procedures described earlier, we estimate the
secondary's spectral type to be M2.5 $\pm$ 1.5, with the secondary alone
having $V = 20.0 \pm 0.4$.  Assuming that the secondary's mass lies in the broad
range from 0.2 to 0.6 M$_{\odot}$, its radius at this $P_{\rm orb}$ would be
0.35 to 0.6 R$_{\odot}$.  Combining this with the surface brightness derived
from the spectral types gives an absolute magnitude $M_V = 10.2 \pm 1.0$.
The distance modulus, uncorrected for extinction, is therefore 
$m - M = +9.8 \pm 1.1$, corresponding to  $910 (+600, -360)$ pc.
\citet{schlegel} estimate a total reddening $E(B-V) = 0.11$ in this
direction; assuming that BI Ori lies beyond all the dust, and taking 
$A_V / E(B-V) = 3.3$ reduces the distance to $\sim 770$ pc.  At maximum
light, BI Ori has $m_p = 13.2$ \citep{gcvs}.  Assuming the color is 
neutral, we find $M_V = 3.4 \pm 1.1$ at maximum.  At BI Ori's period, the 
\citet{warn87} relation predicts
$M_V > 3.6$ (with the brightest value corresponding to $i = 0$).  
This agrees broadly with our nominal value based on the
secondary star's distance, but is a little fainter, 
suggesting that BI Ori is not too far from face-on, or 
a little closer than our nominal distance, or both.  

\subsection{FO Per}

FO Persei was apparently discovered by \citet{morgenroth}, but its 
cataclysmic nature was not immediately recognized.
\citet{fospec} obtained spectra and  
gave equivalent widths for the Balmer lines for
two different nights of observations, between which the continuum changed from
relatively flat to inclined toward the red.  

The emission lines in FO Per are rather narrow (Fig.~\ref{fig:specplot},
Table~\ref{tab:quie}).  This
is often taken to indicate a low orbital inclination.  The 
velocity amplitude $K$ is small,
so that $K/\sigma \approx 1.6 $ for the best fits (Table~\ref{tab:param}).
Because of this, the daily cycle count remains ambiguous; the orbital
frequency is either 5.8 or 6.8 cycle d$^{-1}$, corresponding to 
$P_{\rm orb}$ of 3.52 or 4.13 hr.  CVs with periods in the 3-4 hour
range tend to be novalike variables \citep{shafter92}, whereas FO Per 
is a dwarf nova; thus the 4.13 hr period is more likely {\it a priori}.

\section{Summary}

We have determined the orbital periods of eight CVs without 
significant daily cycle count ambiguity; for FO Per, the period
is narrowed to two choices. For three of the systems
we find high-precision periods by establishing secure cycle 
counts over long baselines.

While most of these objects are similar to others already known, three
stand out as especially interesting.  CZ Aql shows asymmetric, high-velocity 
wings around the Balmer and HeI $\lambda 5876$ and $\lambda 6678$ lines, 
possibly indicating a magnetic system.  BF Eri's proper motion of 
$\sim 100$ mas yr$^{-1}$ is surprising in view of the 
large distance indicated by its secondary spectrum and by the
Warner relation; even if it is somewhat nearer than these indicators
suggest, its kinematics are not typical of disk stars.
Finally, the orbital period of V1006 Cyg places it squarely in the
middle of the so-called period gap between 2 and 3 hours.

{\it Acknowledgments.} 
We are most grateful for support from the National Science Foundation 
through grants AST-9987334 and AST-0307413.  Bill Fenton took
most of the spectra of GZ Cnc, and J. Cameron Brueckner assisted 
with the BF Eri spectroscopy.  Some of the astrometric images of BF Eri
were obtained by S\'ebastien L\'epine and Michael Shara of the 
American Museum of Natural History.  We would like to thank the 
MDM Observatory staff for their skillful and conscientious 
support.  Finally, we are grateful to the Tohono O'odham 
for leasing us their mountain for a while, so that we may 
study the glorious universe in which we all live.

\clearpage

\clearpage

\clearpage

\clearpage

\clearpage

\clearpage

%
\clearpage

%
%
\begin{figure}
\epsscale{0.8}
\caption{Phase-averaged spectra of CZ Aql (top two panels) and 
BF Eri (bottom two panels),
presented as a greyscale.
The scale is inverted, so that emission is represented by darker shades.  The
two CZ Aql spectra are scaled differently to show the line cores (top) and
the extent of the the line wings.   Note the NaD lines in CZ Aql
remain stationary, indicating an interstellar origin.  The feature
at $\lambda 6280$ is telluric.  BF Eri's spectrum is plotted in two
overlapping sections; the K-star's orbital motion is plainly visible. }
\label{fig:trail}
\end{figure}

\clearpage

\begin{deluxetable}{lrrrrrr}
\tabletypesize{\scriptsize}
\tablewidth{0pt}
\tablecolumns{7}
\tablecaption{Stars Observed}
\tablehead{
\colhead{Star} &
\colhead{$\alpha_{2000}$\tablenotemark{a}} &
\colhead{$\delta_{2000}$} &
\colhead{Epoch\tablenotemark{b}} &
\colhead{$V_{\rm obs}$\tablenotemark{c}} & 
\colhead{max\tablenotemark{d}} & 
\colhead{min} \\ 
\colhead{ } &
\colhead{[hh:mm:ss]} &
\colhead{[$^{\circ}$:$'$:$"$]} &
\colhead{} &
\colhead{[mag.]} & 
\colhead{[mag.]} & 
\colhead{[mag.]} \\ 
}
\startdata
LX And & 2:19:44.08 & $+$40:27:22.3 & 2006.7 & 16.3 & 13.5p & 16.4p\\  
CZ Aql &19:19:58.21 & $-$07:10:55.2 & 2003.4 & 15.4 & 13.p  & 15.p \\ 
LU Cam & 5:58:17.86 & $+$67:53:46.2 & 2002.0 & 16.3 & 14.v  & $($16.v \\ 
GZ Cnc & 9:15:51.68 & $+$09:00:49.6 & 2000.3 & 15.4 &  13.1v & 15.4v \\ 
V632 Cyg & 21:36:04.22 & $+$40:26:19.4 & 2000.5 & 17.9 & 12.6p & 17.5p \\  
V1006 Cyg & 19:48:47.20 & $+$57:09:22.8 & 2000.5  & 17.8 & 15.4p & 17.0p \\  
BF Eri &  4:39:29.96 & $-$04:35:59.5 & 2006.2 & 14.8 & 13.5p & 15.5p \\  
BI Ori &  5:23:51.77 & $+$01:00:30.6 & 2002.8 & 17.1 & 13.2p & 16.7p \\  
FO Per &  4:08:34.98 & $+$51:14:48.5 & 2004.0 & 17.1 & 11.8v & 16.v  \\  
\enddata
\tablenotetext{a}{Positions measured from images taken at the 2.4m 
Hiltner telescope, using astrometric solutions from fits to USNO A2.0 \citep{mon96}
or UCAC 2 \citep{ucac2} stars.  Uncertainties are of order 0.1 arcsec.}
\tablenotetext{b}{The date of the image used in the position measurement.  The coordinate
system (equator and equinox) is J2000 in all cases.}
\tablenotetext{c}{Synthesized from spectra, as described in text.}
\tablenotetext{d}{Taken from the GCVS \citep{gcvs}.  Photgraphic magnitudes flagged with
`p', visual with `v'.}
\label{tab:synthmags}
\end{deluxetable}

\clearpage

\begin{deluxetable}{lrccc}
\tabletypesize{\scriptsize}
\tablewidth{0pt}
\tablecolumns{5}
\tablecaption{Journal of Observations}
\tablehead{
\colhead{data} &
\colhead{$N$} &
\colhead{HA start} &
\colhead{HA end} & 
\colhead{telescope} \\
\colhead{(UT)} &
\colhead{ } &
\colhead{[hh:mm]} &
\colhead{[hh:mm]} &
\colhead{ } \\ 
}
\startdata
\cutinhead{LX And} 
2004 Jan 13  &  1  &  +2:35  &  +2:35 & 2.4m\\
2004 Mar 02  &  3  &  +3:45  &  +4:00 & 2.4m\\
2004 Nov 18  &  1  &  +0:29  &  +0:29 & 1.3m\\
2004 Nov 19  &  26  &  $-$3:36  &  +4:17 & 1.3m\\
2004 Nov 19  &  2  &  +2:31  &  +2:35 & 2.4m\\
2004 Nov 20  &  13  &  $-$3:17  &  +3:15 & 1.3m\\
2004 Nov 20  &  1  &  $-$1:23  &  $-$1:23 & 2.4m\\
2006 Jan 19  &  14  &  +1:31  &  +4:10 & 1.3m\\
2006 Jan 22  &  8  &  +3:31  &  +4:46 & 1.3m\\
2007 Jan 26  &  3  &  +1:36  &  +1:58 & 1.3m\\
2007 Jan 27  &  15  &  +0:58  &  +3:39 & 1.3m\\
\cutinhead{CZ Aql} 
2005 Jul 02  &  3  &  +1:40  &  +2:04 & 1.3m\\
2005 Jul 04  &  48  &  $-$3:28  &  +3:17 & 1.3m\\
2005 Jul 05  &  13  &  $-$1:55  &  $-$0:06 & 1.3m\\
2005 Jul 06  &  12  &  $-$3:57  &  +2:28 & 1.3m\\
2005 Sep 03  &  2  &  +1:23  &  +1:32 & 1.3m\\
2005 Sep 07  &  2  &  $-$0:03  &  +0:05 & 1.3m\\
2005 Jun 28  &  2  &  $-$0:01  &  +0:04 & 2.4m\\
2006 Jun 18  &  2  &  +2:05  &  +2:10 & 2.4m\\
2006 Jun 19  &  2  &  +0:36  &  +0:40 & 2.4m\\
2006 Jun 23  &  3  &  $-$1:24  &  $-$1:12 & 2.4m\\
\tablebreak
\cutinhead{LU Cam}
2002 Jan 22  &  2  &  $-$1:40  &  $-$1:19 & 2.4m\\
2002 Jan 23  &  8  &  $-$1:33  &  +2:31 & 2.4m\\
2002 Jan 24  &  25  &  $-$3:15  &  +5:47 & 2.4m\\
2002 Feb 18  &  2  &  +2:20  &  +2:28 & 2.4m\\
2002 Feb 19  &  2  &  +2:41  &  +2:50 & 2.4m\\
2002 Feb 20  &  4  &  $-$0:02  &  +3:43 & 2.4m\\
2002 Feb 22  &  1  &  +2:18  &  +2:18 & 2.4m\\
2004 Jan 16  &  2  &  +2:49  &  +2:53 & 2.4m\\
2004 Jan 17  &  1  &  +0:40  &  +0:40 & 2.4m\\
2004 Jan 19  &  1  &  $-$0:32  &  $-$0:32 & 2.4m\\
2004 Mar 02  &  1  &  +0:53  &  +0:53 & 2.4m\\
2004 Mar 07  &  6  &  +2:29  &  +3:14 & 2.4m\\
2004 Nov 19  &  4  &  +2:48  &  +3:27 & 2.4m\\
2005 Mar 21  &  1  &  +1:12  &  +1:12 & 2.4m\\
2005 Mar 22  &  2  &  +1:21  &  +1:30 & 2.4m\\
2005 Sep 09  &  1  &  $-$2:00  &  $-$2:00 & 2.4m\\
2005 Sep 12  &  1  &  $-$1:49  &  $-$1:49 & 2.4m\\
2006 Jan 09  &  2  &  +1:49  &  +1:55 & 2.4m\\
\cutinhead{GZ Cnc}
2000 Apr 07  &  2  &  +4:26  &  +4:32 & 2.4m\\
2000 Apr 08  &  1  &  +0:27  &  +0:27 & 2.4m\\
2000 Apr 10  &  5  &  $-$0:31  &  +4:32 & 2.4m\\
2000 Apr 11  &  15  &  +2:19  &  +4:22 & 2.4m\\
2001 Mar 24  &  2  &  +4:39  &  +4:49 & 2.4m\\
2001 Mar 25  &  21  &  $-$1:10  &  +3:10 & 2.4m\\
2001 Mar 26  &  20  &  $-$0:50  &  +4:27 & 2.4m\\
2001 Mar 27  &  2  &  $-$0:08  &  +0:01 & 2.4m\\
2001 Mar 28  &  3  &  +0:14  &  +0:25 & 2.4m\\
\cutinhead{V632 Cyg}
2005 Jul 07  &  2  &  +1:00  &  +1:16 & 1.3m\\
2005 Jul 08  &  10  &  $-$5:12  &  +0:33 & 1.3m\\
2005 Jul 09  &  18  &  $-$5:03  &  +1:09 & 1.3m\\
2005 Jul 10  &  18  &  $-$5:09  &  +1:07 & 1.3m\\
2005 Jul 11  &  3  &  +0:52  &  +1:19 & 1.3m\\
\tablebreak
\cutinhead{V1006 Cyg}
2003 Jun 22  &  1  &  +1:06  &  +1:06 & 2.4m\\
2004 Jun 24  &  5  &  +0:52  &  +1:56 & 1.3m\\
2004 Jun 25  &  5  &  $-$1:43  &  +0:34 & 1.3m\\
2004 Jun 25  &  1  &  +4:06  &  +4:06 & 2.4m\\
2004 Jun 26  &  10  &  $-$4:25  &  +1:56 & 1.3m\\
2004 Jun 27  &  3  &  $-$3:00  &  $-$2:00 & 1.3m\\
2004 Jun 28  &  4  &  $-$2:25  &  +1:02 & 1.3m\\
2004 Jun 28  &  1  &  +0:28  &  +0:28 & 2.4m\\
2004 Jun 29  &  5  &  +0:55  &  +1:59 & 1.3m\\
2004 Jun 29  &  1  &  +4:07  &  +4:07 & 2.4m\\
2004 Jun 30  &  18  &  $-$4:39  &  +2:26 & 1.3m\\
2004 Jul 01  &  10  &  $-$3:58  &  +1:53 & 2.4m\\
2005 Jul 05  &  13  &  +0:58  &  +3:11 & 1.3m\\
2004 Jun 30  &  4  &  $-$3:41  &  +3:52 & 2.4m\\
2004 Jul 01  &  12  &  $-$4:07  &  +2:17 & 1.3m\\
2005 Jul 03  &  27  &  $-$3:57  &  +1:20 & 1.3m\\
2005 Jul 05  &  13  &  +0:58  &  +3:11 & 1.3m\\
\cutinhead{BF Eri}
2001 Dec 18  &  3  &  +2:35  &  +2:56 & 1.3m\\
2001 Dec 19  &  10  &  $-$3:06  &  +4:04 & 1.3m\\
2001 Dec 20  &  12  &  $-$3:49  &  +2:00 & 1.3m\\
2001 Dec 21  &  2  &  +3:04  &  +3:14 & 1.3m\\
2001 Dec 22  &  5  &  $-$3:13  &  $-$2:31 & 1.3m\\
2001 Dec 23  &  12  &  $-$3:18  &  +4:35 & 1.3m\\
2001 Dec 24  &  13  &  $-$3:17  &  +3:08 & 1.3m\\
2001 Dec 25  &  14  &  $-$2:20  &  +1:07 & 1.3m\\
2001 Dec 26  &  8  &  $-$2:10  &  +3:05 & 1.3m\\
2001 Dec 27  &  18  &  $-$2:39  &  +4:14 & 1.3m\\
2002 Jan 19  &  1  &  +1:27  &  +1:27 & 2.4m\\
2002 Jan 20  &  2  &  $-$1:33  &  +2:13 & 2.4m\\
2002 Jan 22  &  1  &  $-$2:02  &  $-$2:02 & 2.4m\\
2002 Feb 21  &  2  &  +1:26  &  +1:35 & 2.4m\\
2002 Feb 22  &  2  &  +0:40  &  +0:49 & 2.4m\\
2002 Oct 26  &  2  &  $-$0:11  &  +0:05 & 2.4m\\
2002 Oct 31  &  1  &  +3:21  &  +3:21 & 2.4m\\
2003 Feb 02  &  1  &  +0:04  &  +0:04 & 2.4m\\
2005 Sep 11  &  2  &  $-$0:57  &  $-$0:40 & 1.3m\\
2006 Mar 16  &  1  &  +1:56  &  +1:56 & 1.3m\\
2006 Mar 17  &  5  &  +2:15  &  +2:57 & 1.3m\\
2007 Jan 28  &  9  &  $-$1:46  &  $-$0:13 & 1.3m\\
\cutinhead{BI Ori}
2006 Jan 20  &  32  &  $-$2:54  &  +4:05 & 1.3m\\
2006 Jan 21  &  22  &  $-$2:47  &  +2:21 & 1.3m\\
2006 Jan 23  &  6  &  +1:03  &  +2:08 & 1.3m\\
\cutinhead{FO Per}
1995 Oct 09  &  9  &  $-$4:40  &  $-$3:26 & 2.4m\\
1995 Oct 10  &  5  &  $-$5:22  &  +1:26 & 2.4m\\
1996 Dec 19  &  14  &  +1:38  &  +4:03 & 1.3m\\
1996 Dec 20  &  5  &  +2:08  &  +3:43 & 1.3m\\
2001 Dec 18  &  9  &  $-$2:35  &  +4:36 & 1.3m\\
2004 Nov 18  &  1  &  $-$0:11  &  $-$0:11 & 1.3m\\
2004 Nov 19  &  9  &  +2:52  &  +5:08 & 1.3m\\
2004 Nov 19  &  2  &  +3:24  &  +2:48 & 2.4m\\
2004 Nov 20  &  20  &  $-$0:56  &  +5:10 & 1.3m\\
2006 Jan 10  &  12  &  $-$0:35  &  +2:35 & 1.3m\\
2006 Jan 10  &  1  &  +4:21  &  +4:21 & 2.4m\\
2006 Jan 11  &  18  &  $-$1:29  &  +2:41 & 1.3m\\
2006 Jan 11  &  1  &  $-$3:12  &  $-$3:12 & 2.4m\\
2006 Jan 12  &  4  &  $-$1:16  &  $-$0:36 & 1.3m\\
2006 Jan 13  &  10  &  +3:44  &  +5:37 & 1.3m\\
2006 Jan 16  &  11  &  $-$1:36  &  +1:00 & 1.3m\\
\enddata
\label{tab:obsjournal}
\end{deluxetable}

\clearpage

\begin{deluxetable}{lrcc}
\tabletypesize{\scriptsize}
\tablewidth{0pt}
\tablecolumns{4}
\tablecaption{Spectral Features in Quiescence}
\tablehead{
&
\colhead{E.W.\tablenotemark{a}} &
\colhead{Flux}  &
\colhead{FWHM \tablenotemark{b}} \\
\colhead{Feature} &
\colhead{(\AA )} &
\colhead{(10$^{-16}$ erg cm$^{-2}$ s$^{1}$)} &
\colhead{(\AA)} \\
}
\startdata
\cutinhead{LX And}
            H$\beta$ & $ 45$ & $690$ & 18 \\
  HeI $\lambda 4921$ & $  4$ & $ 60$ & 25 \\
  HeI $\lambda 5015$ & $  3$ & $ 50$ & 20 \\
   Fe $\lambda 5169$ & $  2$ & $ 20$ & 14 \\
  HeI $\lambda 5876$ & $ 11$ & $120$ & 19 \\
           H$\alpha$ & $ 54$ & $560$ & 17 \\
  HeI $\lambda 6678$ & $  4$ & $ 40$ & 19 \\
\cutinhead{CZ Aql}
            H$\beta$ & $ 21$ & $670$ & 18 \\
  HeI $\lambda 4921$ & $  1$ & $ 40$ & 12 \\
  HeI $\lambda 5015$ & $  2$ & $ 50$ & 13 \\
  HeI $\lambda 5876$ & $  7$ & $160$ & 15 \\
                 NaD & $ -1$ & $-16$ & \nodata \\
           H$\alpha$ & $ 61$ & $1250$ & 27 \\
  HeI $\lambda 6678$ & $  5$ & $ 90$ & 16 \\
\cutinhead{LU Cam}
           H$\gamma$ & $ 10$ & $170$ & 12 \\
  HeI $\lambda 4471$ & $  2$ & $ 30$ & 10 \\
            H$\beta$ & $ 14$ & $190$ & 12 \\
  HeI $\lambda 4921$ & $  1$ & $ 20$ & 12 \\
  HeI $\lambda 5015$ & $  2$ & $ 20$ & 13 \\
   Fe $\lambda 5169$ & $  1$ & $ 10$ & 12 \\
  HeI $\lambda 5876$ & $  4$ & $ 50$ & 12 \\
           H$\alpha$ & $ 27$ & $240$ & 13 \\
  HeI $\lambda 6678$ & $  3$ & $ 20$ & 15 \\
  HeI $\lambda 7067$ & $  3$ & $ 20$ & \nodata \\
\cutinhead{GZ Cnc}
           H$\gamma$ & $ 26$ & $940$ & 26 \\
  HeI $\lambda 4471$ & $  8$ & $260$ & 28 \\
 HeII $\lambda 4686$ & $  5$ & $140$ & 46 \\
            H$\beta$ & $ 36$ & $1040$ & 25 \\
  HeI $\lambda 4921$ & $  5$ & $140$ & 26 \\
  HeI $\lambda 5015$ & $  4$ & $100$ & 28 \\
   Fe $\lambda 5169$ & $  2$ & $ 60$ & 26 \\
  HeI $\lambda 5876$ & $  9$ & $200$ & 27 \\
           H$\alpha$ & $ 38$ & $790$ & 25 \\
  HeI $\lambda 6678$ & $  4$ & $ 90$ & 31 \\
  HeI $\lambda 7067$ & $  3$ & $ 60$ & 32 \\
\cutinhead{V632 Cyg}
            H$\beta$ & $ 80$ & $260$ & 24 \\
  HeI $\lambda 4921$ & $  6$ & $ 20$ & 27 \\
  HeI $\lambda 5015$ & $  8$ & $ 20$ & 26 \\
   Fe $\lambda 5169$ & $  5$ & $ 10$ & 26 \\
  HeI $\lambda 5876$ & $ 28$ & $ 70$ & 27 \\
           H$\alpha$ & $113$ & $260$ & 27 \\
  HeI $\lambda 6678$ & $ 15$ & $ 30$ & 32 \\
\cutinhead{V1006 Cyg}
            H$\beta$ & $ 74$ & $250$ & 27 \\
  HeI $\lambda 4921$ & $  8$ & $ 30$ & 30 \\
  HeI $\lambda 5015$ & $  8$ & $ 20$ & 28 \\
   Fe $\lambda 5169$ & $  8$ & $ 30$ & 28 \\
  HeI $\lambda 5876$ & $ 26$ & $ 70$ & 34 \\
           H$\alpha$ & $108$ & $250$ & 31 \\
  HeI $\lambda 6678$ & $ 11$ & $ 30$ & 38 \\
\cutinhead{BF Eri}
 HeII $\lambda 4686$ & $  6$ & $240$ & 46 \\
            H$\beta$ & $ 23$ & $1060$ & 22 \\
  HeI $\lambda 5015$ & $  2$ & $110$ & 29 \\
  HeI $\lambda 5876$ & $  5$ & $240$ & 21 \\
           H$\alpha$ & $ 27$ & $1200$ & 22 \\
  HeI $\lambda 6678$ & $  3$ & $120$ & 27 \\
\cutinhead{BI Ori}
            H$\beta$ & $ 34$ & $180$ & 32 \\
  HeI $\lambda 4921$ & $  6$ & $ 30$ & 36 \\
  HeI $\lambda 5015$ & $  6$ & $ 30$ & 43 \\
   Fe $\lambda 5169$ & $  5$ & $ 30$ & 39 \\
  HeI $\lambda 5876$ & $  6$ & $ 30$ & 31 \\
           H$\alpha$ & $ 36$ & $190$ & 31 \\
\cutinhead{FO Per}
            H$\beta$ & $ 24$ & $160$ &  9 \\
  HeI $\lambda 4921$ & $  2$ & $ 20$ &  7 \\
  HeI $\lambda 5015$ & $  3$ & $ 20$ &  7 \\
   Fe $\lambda 5169$ & $  2$ & $ 10$ &  7 \\
  HeI $\lambda 5876$ & $  5$ & $ 30$ &  7 \\
           H$\alpha$ & $ 29$ & $170$ &  9 \\
  HeI $\lambda 6678$ & $  2$ & $ 10$ &  9 \\
\label{tab:quie}
\enddata

\tablenotetext{a}{Emission equivalent widths are counted as positive.}
\tablenotetext{b}{From Gaussian fits.}

\end{deluxetable}

\clearpage

\begin{deluxetable}{llrrrr}
\tabletypesize{\scriptsize}
\tablewidth{0pt}
\tablecolumns{6}
\tablecaption{Radial Velocities}
\tablehead{
\colhead{Star} &
\colhead{time \tablenotemark{a}} &
\colhead{$v_{\rm abs}$} &
\colhead{$\sigma_{v_{\rm abs}}$} &
\colhead{$v_{\rm emn}$} &
\colhead{$\sigma_{v_{\rm emn}}$} \\
&
&
\colhead{(km s$^{-1}$)} &
\colhead{(km s$^{-1}$)} &
\colhead{(km s$^{-1}$)} &
\colhead{(km s$^{-1}$)} \\
 }
\startdata
LX And & 53017.7061 &  \nodata  &  \nodata  &   $-$44 &  $-$11 \\
LX And & 53066.6160 &  \nodata  &  \nodata  &   $-$57 &   $-$8 \\
LX And & 53066.6208 &  \nodata  &  \nodata  &   $-$37 &   $-$8 \\
LX And & 53066.6263 &  \nodata  &  \nodata  &   $-$77 &   $-$7 \\
\enddata
\tablenotetext{a}{Heliocentric Julian date of mid-integration,
minus 2400000.}
\tablecomments{All emission-line velocities are of H$\alpha$.  
Emission-line velocity uncertainties are derived from 
counting statistics and should be regarded as lower limits. 
Table \ref{tab:vels} is published in its entirety in the 
electronic version of the Publications of the Astronomical Society
of the Pacific.  A short portion is shown here for guidance regarding its
form and content.}
\label{tab:vels}
\end{deluxetable}

\clearpage

\begin{deluxetable}{llllrrcc}
\tablecolumns{8}
\tabletypesize{\scriptsize}
\tablewidth{0pt}
\tablecaption{Fits to Radial Velocities}
\tablehead{
\colhead{Star} &
\colhead{Algorithm\tablenotemark{a}} &
\colhead{$T_0$\tablenotemark{b}} &
\colhead{$P$} &
\colhead{$K$} &
\colhead{$\gamma$} &
\colhead{$N$} &
\colhead{$\sigma$\tablenotemark{c}}  \\
&
&
&
\colhead{(d)} &
\colhead{(km s$^{-1}$)} &
\colhead{(km s$^{-1}$)} &
&
\colhead{(km s$^{-1}$)} \\
}
\startdata
LX And & G2,21,7 & 53754.6861(12) & 0.1509743(5) &  81(4) & $-48(3)$ & 87 &  15 \\ 
CZ Aql & G2,18,8 & 53557.880(2) & 0.2005(6)\tablenotemark{d} &  193(15) & $ 5(10)$ & 89 &  61 \\
LU Cam & D,15 & 52327.7421(14) & 0.1499686(4) &  57(4) & $ 44(3)$ & 66 &  14 \\
GZ Cnc & G2,15,9 & 51992.8928(13) & 0.0881(4)\tablenotemark{d}  &  79(7) & $ 22(5)$ & 71 &  26 \\
V632 Cyg & D,28  & 53560.9746(13) & 0.06377(8) &  62(8) & $-49(5)$ & 51 &  28 \\
V1006 Cyg & G2,20,9  & 53187.9091(16) & 0.09904(9)\tablenotemark{d}  &  89(8) & $-11(6)$ & 120 &  44 \\
BF Eri emission & G2,21,7 & 52574.0027(18) & 0.2708801(6) &  109(5) & $-91(3)$ & 126 &  24 \\
BF Eri absorption & \nodata  & 52573.8632(9) & 0.2708805(4) &  182(4) & $-72(3)$ & 117 &  20 \\
BF Eri mean:      &  \nodata &  \nodata   &  0.2708804(4) & \nodata &\nodata  &\nodata  & \\
BI Ori & 53756.541(3) & G2,35,9 & 0.1915(5) &  131(13) & $ 24(9)$ & 60 &  44 \\
FO Per (shorter) & D,11  & 52261.872(3) & 0.1467(4)\tablenotemark{d}  &  27(3) & $-49(2)$ & 131 &  17 \\
FO Per (longer) & D,11 & 52261.893(3) & 0.1719(5)\tablenotemark{d}  &  27(3) & $-45(2)$ & 131 &  17 \\
\enddata
\tablecomments{Parameters of sinusoidal least-squares fits to the velocity timeseries, 
of the form $v(t) = \gamma + K \sin(2 \pi (t - T_0) / P)$.  The quoted parameter uncertainties 
are based on the assumption that the scatter of the data around the best fit is a 
realistic estimate of the velocity uncertainty \citep{cash}.  In practice this is more conservative
than assuming that counting statistics uncertainties are realistic.}
\tablenotetext{a}{Code for the convolution function used to derive emission line velocities;
D = derivative of a Gaussian, G2 = double-Gaussian function (see text).  For the D
algorithm the number that follows gives the line full-width at half-maximum, in \AA ,
for which the function is optimized; for the G2 algorithm the two numbers are respectively
the separation of the two Gaussians and their individual FWHMa, again in \AA .}
\tablenotetext{b}{Heliocentric Julian Date minus 2400000.  The epoch is chosen
to be near the center of the time interval covered by the data, and
within one cycle of an actual observation.}
\tablenotetext{c}{Root-mean-square residual of the fit.}
\tablenotetext{d}{The period determination in this case is complicated by unknown numbers
of cycles between observing runs; the uncertainty given here is an estimate based on 
fits to individual runs.  Only certain values within the period range given here are 
allowed; see text for details.}
\label{tab:param}
\end{deluxetable}



\clearpage

\begin{thebibliography}


\bibitem[Andronov \& Pinsonneault(2004)]{andronov} Andronov, N., 
\& Pinsonneault, M.~H.\ 2004, \apj, 614, 326 

\bibitem[Araujo-Betancor et al.(2005)]{hasitall} 
Araujo-Betancor, S., et al.\ 2005, \aap, 430, 629 

\bibitem[Baraffe \& Kolb(2000)]{baraffe} Baraffe, I., \& Kolb, 
U.\ 2000, \mnras, 318, 354 

\bibitem[Bessell(1990)]{bessell} Bessell, M.~S.\ 1990, \pasp, 102, 1181


\bibitem[Beuermann(2006)]{beuermann06} Beuermann, K.\ 2006, \aap, 
460, 78

\bibitem[Boeshaar(1976)]{boeshaar} Boeshaar, P. 1976, Ph. D. thesis, Ohio State
University

\bibitem[Bruch(1989)]{fospec} Bruch, A.\ 1989, \aaps, 78, 145

\bibitem[Bruch \& Schimpke(1992)]{v1006spec} Bruch, A., \& 
Schimpke, T.\ 1992, \aaps, 93, 419

\bibitem[Casares et al.(1996)]{v795her} Casares, J., 
Martinez-Pais, I.~G., Marsh, T.~R., Charles, P.~A., \& Lazaro, C.\ 1996, 
\mnras, 278, 219 

\bibitem[Cash(1979)]{cash} Cash, W.\ 1979, \apj, 228, 939 


\bibitem[Cieslinski et al.(1998)]{ciess} Cieslinski, D.,
Steiner, J.~E., \& Jablonski, F.~J.\ 1998, \aaps, 131, 119 

\bibitem[Dickinson et al.(1997)]{v795her2} Dickinson, R.~J., 
Prinja, R.~K., Rosen, S.~R., King, A.~R., Hellier, C., \& Horne, K.\ 1997, 
\mnras, 286, 447 

\bibitem[Downes et al.(2001)]{Downes} Downes, R.~A., Webbink,
R.~F., Shara, M.~M., Ritter, H., Kolb, U., \& Duerbeck, H.~W.\ 2001, \pasp,
113, 764

\bibitem[Elvis et al.(1992)]{slew} Elvis, M., Plummer, D., 
Schachter, J., \& Fabbiano, G.\ 1992, \apjs, 80, 257 

\bibitem[Hellier \& Naylor(1998)]{helliergap} Hellier, C., \& 
Naylor, T.\ 1998, \mnras, 295, L50 


\bibitem[Jenniskens \& Desert(1994)]{dib} Jenniskens, P., 
\& Desert, F.-X.\ 1994, \aaps, 106, 39

\bibitem[Jiang et al.(2000)]{jiang} Jiang, X.~J., Engels, D., 
Wei, J.~Y., Tesch, F., \& Hu, J.~Y.\ 2000, \aap, 362, 263 

\bibitem[Kato(1999)]{kat99} Kato, T.\ 1999, Informational 
Bulletin on Variable Stars, 4745, 1 


\bibitem[Kato et al.(2002)]{kat02} Kato, T., et al.\ 2002,
\aap, 396, 929

\bibitem[Keenan \& McNeil(1989)]{keenan} Keenan, P.~C., \& 
McNeil, R.~C.\ 1989, \apjs, 71, 245 

\bibitem[Kholopov et al.(1999)]{gcvs} Kholopov, P.~N., et 
al.\ 1999, VizieR Online Data Catalog, 2214, 0 

\bibitem[Kinman et al.(1982)]{lick82} Kinman, T.~D., Mahaffey, 
C.~T., \& Wirtanen, C.~A.\ 1982, \aj, 87, 314

\bibitem[Hanson et al.(2004)]{hanson} Hanson, R.~B., Klemola, 
A.~R., Jones, B.~F., \& Monet, D.~G.\ 2004, \aj, 128, 1430

\bibitem[Kurtz \& Mink(1998)]{kurtzmink} Kurtz, M.~J., \& Mink, 
D.~J.\ 1998, \pasp, 110, 934 

\bibitem[Liu et al.(1999)]{v632spec} Liu, W., Hu, J.~Y., Zhu,
X.~H., \& Li, Z.~Y.\ 1999, \apjs, 122, 243

\bibitem[Marsh(1988)]{marsh88} Marsh, T.~R.\ 1988, \mnras, 231, 
1117 

\bibitem[Monet et al.(1996)]{mon96} Monet, D. et al. 1996,
USNO-A2.0, (U. S. Naval Observatory, Washington, DC)

\bibitem[Morgenroth(1939)]{morgenroth} Morgenroth, O.\ 1939, 
Astronomische Nachrichten, 268, 273 
 

\bibitem[Morales-Rueda \& Marsh(2002)]{spec} Morales-Rueda, 
L., \& Marsh, T.~R.\ 2002, \mnras, 332, 814 

\bibitem[Patterson et al.(2002)]{rx} Patterson, J., et 
al.\ 2002, \pasp, 114, 1364 

\bibitem[Robinson(1992)]{robinson} Robinson, E.~L.\ 1992, ASP 
Conf.~Ser.~ 29: Cataclysmic Variable Stars, 29, 3 

\bibitem[Schachter et al.(1996)]{bfspec} Schachter, J.~F., 
Remillard, R., Saar, S.~H., Favata, F., Sciortino, S., \& Barbera, M.\ 
1996, \apj, 463, 747 

\bibitem[Schlegel et al.(1998)]{schlegel} Schlegel, D.~J., 
Finkbeiner, D.~P., \& Davis, M.\ 1998, \apj, 500, 525 

\bibitem[Schneider \& Young(1980)]{schyo} Schneider, D.~P.~\&
Young, P.\ 1980, \apj, 238, 946

\bibitem[Shafter(1983)]{shaf} Shafter, A.~W.\ 1983, \apj, 267, 222

\bibitem[Shafter(1992)]{shafter92} Shafter, A.~W.\ 1992, \apj, 
394, 268


\bibitem[Szkody(1987)]{biori87} Szkody, P.\ 1987, \apjs, 63, 
685 

\bibitem[Tappert \& Bianchini(2003)]{tappert} Tappert, C., \& 
Bianchini, A.\ 2003, \aap, 401, 1101 

\bibitem[Taylor et al.(1999)]{lspeg} Taylor, C.~J., 
Thorstensen, J.~R., \& Patterson, J.\ 1999, \pasp, 111, 184 

\bibitem[Thorstensen(2003)]{parallax} Thorstensen, J.~R.\ 2003, 
\aj, 126, 3017 

\bibitem[Thorstensen et al.(2004)]{longp03} Thorstensen, J.~R., 
Fenton, W.~H., \& Taylor, C.~J.\ 2004, \pasp, 116, 300 

\bibitem[Thorstensen \& Freed(1985)]{tf85} Thorstensen,
J.~R.~\& Freed, I.~W.\ 1985, \aj, 90, 2082

\bibitem[Thorstensen et al.(1996)]{tpst} Thorstensen, J.~R., Patterson,
J.~O., Shambrook, A., \& Thomas, G.\ 1996, \pasp, 108, 73

\bibitem[Thorstensen et al.(1991)]{thorsw} Thorstensen, J.~R., 
Ringwald, F.~A., Wade, R.~A., Schmidt, G.~D., \& Norsworthy, J.~E.\ 1991, 
\aj, 102, 272 

\bibitem[Thorstensen \& Taylor(2000)]{v533her} Thorstensen, 
J.~R., \& Taylor, C.~J.\ 2000, \mnras, 312, 629 

\bibitem[Uemura et al.(2000)]{jan00} Uemura, M., Kato, T., \& 
Watanabe, M.\ 2000, Informational Bulletin on Variable Stars, 4831, 1 

\bibitem[Warner(1987)]{warn87} Warner, B.\ 1987, \mnras, 227, 23

\bibitem[Warner(1995)]{warner} Warner, B.\ 1995, Cambridge 
Astrophysics Series, Cambridge, New York: Cambridge University Press, 
|c1995  

\bibitem[Zacharias et al.(2004)]{ucac2} Zacharias, N., Urban, 
S.~E., Zacharias, M.~I., Wycoff, G.~L., Hall, D.~M., Monet, D.~G., \& 
Rafferty, T.~J.\ 2004, \aj, 127, 3043 

\end{thebibliography}
\end{document}